\documentclass[aps,prl,onecolumn,superscriptaddress]{revtex4}
\usepackage{graphicx}

\begin{document}

\title{Nature of magnetic excitations in superconducting BaFe$_{1.9}$Ni$_{0.1}$As$_{2}$}
\author{Mengshu Liu}
\affiliation{
Department of Physics and Astronomy, The University of Tennessee, Knoxville, Tennessee 37996-1200, USA
}
\author{Leland W. Harriger}
\affiliation{
Department of Physics and Astronomy, The University of Tennessee, Knoxville, Tennessee 37996-1200, USA
}
\author{Huiqian Luo}
\affiliation{Beijing National Laboratory for Condensed Matter Physics,
Institute of Physics, Chinese Academy of Sciences, Beijing 100190, China
}
\author{Meng Wang}
\affiliation{Beijing National Laboratory for Condensed Matter Physics,
Institute of Physics, Chinese Academy of Sciences, Beijing 100190, China
}
\affiliation{
Department of Physics and Astronomy, The University of Tennessee, Knoxville, Tennessee 37996-1200, USA
}
\author{R. A. Ewings}
\affiliation{
ISIS Facility, Rutherford Appleton Laboratory, Chilton, Didcot, Oxfordshire OX11 0QX, UK
}
\author{T. Guidi}
\affiliation{
ISIS Facility, Rutherford Appleton Laboratory, Chilton, Didcot, Oxfordshire OX11 0QX, UK
}
\author{Hyowon Park}
\affiliation{
Department of Physics, Rutgers University, Piscataway, NJ 08854, USA
}
\author{Kristjan Haule}
\affiliation{
Department of Physics, Rutgers University, Piscataway, NJ 08854, USA
}
\author{Gabriel Kotliar}
\affiliation{
Department of Physics, Rutgers University, Piscataway, NJ 08854, USA
}
\author{S. M. Hayden}
\affiliation{
H. H. Wills Physics Laboratory, University of Bristol, Tyndall Avenue,      
Bristol, BS8 1TL, UK 
}
\author{Pengcheng Dai}
\affiliation{
Department of Physics and Astronomy, The University of Tennessee, Knoxville, Tennessee 37996-1200, USA
}
\affiliation{Beijing National Laboratory for Condensed Matter Physics,
Institute of Physics, Chinese Academy of Sciences, Beijing 100190, China
}


\maketitle
{\bf Since the discovery of the metallic antiferromagnetic
  (AF) ground state near superconductivity in iron-pnictide
  superconductors \cite{kamihara,cruz,rickgreen}, a central question
  has been whether magnetism in these materials arises from weakly correlated
  electrons~\cite{mazin,jdong}, as in the case of spin-density-wave in
  pure chromium \cite{fawcett}, requires strong electron
  correlations~\cite{HG}, or can even be described in terms of 
  localized electrons~\cite{si,xu} such as the AF insulating state of copper
  oxides \cite{palee}.
  Here we use inelastic neutron
  scattering to determine the absolute intensity of the magnetic excitations
  throughout the Brillouin zone in electron-doped superconducting
  BaFe$_{1.9}$Ni$_{0.1}$As$_{2}$ ($T_c=20$ K), which allows us to obtain the size of
  the fluctuating magnetic moment $\left\langle m^2\right\rangle$, and its energy distribution \cite{inosov,lester10}.
  We find that superconducting BaFe$_{1.9}$Ni$_{0.1}$As$_{2}$ and AF 
  BaFe$_2$As$_2$ \cite{harriger} both have fluctuating magnetic moments $\left\langle m^2\right\rangle\approx3.2\ \mu_B^2$ per Fe(Ni), 
  which are similar to those found 
  in the AF insulating copper oxides \cite{headings,hayden}.  The common
  theme in both classes of high temperature superconductors is that
  magnetic excitations have partly localized character, thus
  showing the importance of strong correlations for high temperature
  superconductivity \cite{bosov}.
 }

In the undoped state, iron pnictides such as BaFe$_{2}$As$_{2}$ form a
metallic low-temperature orthorhombic phase with the AF structure as shown
in Fig. 1a \cite{huang}.  Inelastic neutron scattering measurements
have mapped out spin waves throughout the Brioullion zone in the AF
orthorhombic and paramagnetic tetragonal phases \cite{harriger}.  Upon
Co- and Ni-doping to induce optimal superconductivity via electron doping, the
orthorhombic structural distortion and static AF order in
BaFe$_{2}$As$_{2}$ are suppressed and the system becomes tetragonal
and paramagnetic at all temperatures \cite{lester}.  In previous
inelastic neutron scattering experiments on optimally electron-doped
Ba(Fe,Co,Ni)$_2$As$_2$ superconductors
\cite{inosov,lester10,lumsden,chi,hfli,mywang10}, spin excitations up
to $\sim$120 meV were observed.  However, the lack of spin excitations
data at higher energies in absolute units precluded a 
comparison with spin waves in undoped BaFe$_2$As$_2$.  Only the
absolute intensity measurements in the entire Brillouin zone
can reveal the effect of electron-doping on the overall spin
excitations spectra and allow a direct comparison with the results in
the AF insulating copper oxides~\cite{headings,hayden}.  For the experiments, we chose to study well-characterized electron-doped 
BaFe$_{1.9}$Ni$_{0.1}$As$_{2}$ \cite{chi,mywang10} because large single crystals were available \cite{yanchao}
and their properties are similar to Co-doped BaFe$_2$As$_2$ \cite{inosov,lester10,lumsden,hfli,budko}.

By comparing spin excitations in BaFe$_{1.9}$Ni$_{0.1}$As$_{2}$ and BaFe$_2$As$_2$
throughout the Brillouin zone, we were able to probe how electron-doping and superconductivity affect the overall 
spin excitations spectra. We demonstrate that while the low-energy spin excitations are affected, 
the high-energy excitations show a very weak temperature and doping dependence.  
Comparison of our results with various theories suggests that neither a fully itinerant nor a localized picture explains the magnetic excitation spectrum.  However, a combination of density functional theory (DFT) and dynamic mean field theory (DMFT) provides a natural way to improve on both these pictures.

\begin{figure}[t]
\includegraphics[scale=.37]{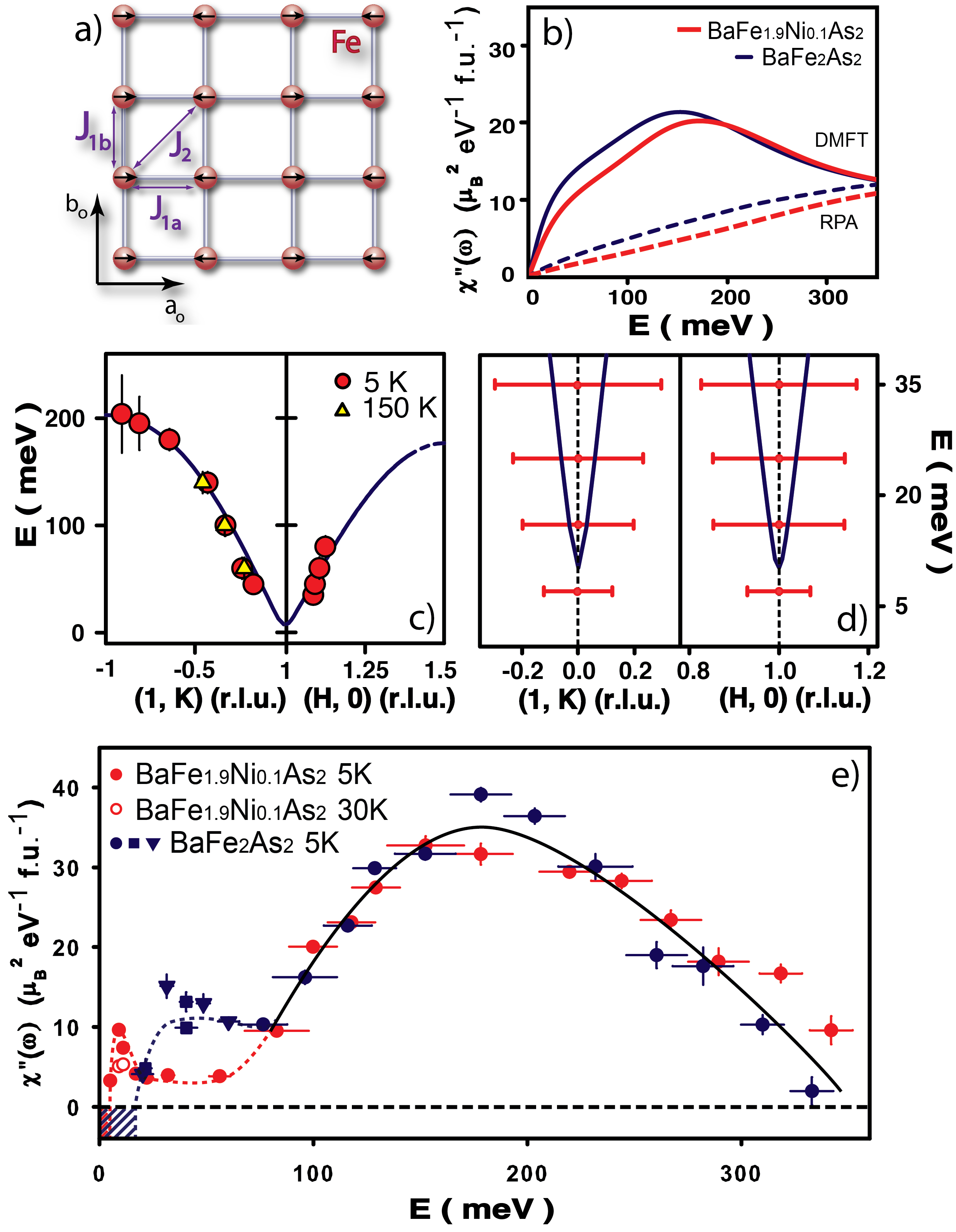}
\caption{ Summary of neutron scattering and calculation results.
Our experiments were carried out on the MERLIN time-of-flight chopper spectrometer at the Rutherford-Appleton Laboratory, UK \cite{merlin}.
We co-aligned 28 g of single crystals of BaFe$_{1.9}$Ni$_{0.1}$As$_{2}$ (with in-plane mosaic of 2.5$^\circ$ and
out-of-plane mosaic of 4$^\circ$). The incident beam energies were $E_i= 20, 25, 30, 80, 250, 450, 600$ meV, and mostly with $E_i$ parallel to the $c$-axis. To facilitate easy comparison with spin waves in BaFe$_{2}$As$_{2}$ \cite{harriger}, we
defined the wave vector $Q$ at ($q_x$, $q_y$, $q_z$) as $(H,K,L) = (q_xa/2\pi, q_yb/2\pi, q_zc/2\pi)$ reciprocal lattice units (rlu) using the
orthorhombic unit cell, where $a = b = 5.564$ \AA, and $c = 12.77$ \AA.  The data are normalized to absolute units using a vanadium standard \cite{harriger}, which may have a systematic error up to 20\% due to differences in neutron 
illumination of vanadium and sample, and time-of-flight instruments.
(a) AF spin structure of
 BaFe$_{2}$As$_{2}$ with Fe spin ordering.  The effective magnetic exchange couplings along different directions
 are depicted. (b) RPA and LDA+DMFT calculations of $\chi^{\prime\prime}(\omega)$ in absolute units for
 BaFe$_2$As$_2$ and BaFe$_{1.9}$Ni$_{0.1}$As$_2$.
(c) The solid lines show spin wave dispersions of BaFe$_2$As$_2$ for
$J_{1a}\neq J_{1b}$ along the $[1,K]$ and $[H,0]$ directions
obtained in Ref.~\cite{harriger}.  The filled circles and upper triangles
are spin excitation dispersions of BaFe$_{1.9}$Ni$_{0.1}$As$_{2}$ at 5 K and 150 K, respectively.
 (d) The solid line shows low energy spin waves of BaFe$_{2}$As$_{2}$.  The horizontal bars show the full-width-half-maximum
of spin excitations in BaFe$_{1.9}$Ni$_{0.1}$As$_{2}$.
(e) Energy dependence of $\chi^{\prime\prime}(\omega)$ for BaFe$_2$As$_2$ (filled blue circles)
and BaFe$_{1.9}$Ni$_{0.1}$As$_2$ below (filled red circles) and above (open red circles) $T_c$.
The solid and dashed lines are guide to the line. The vertical error
bars indicate the statistical errors of one standard deviation.  The horizontal error bars in (e) 
indicate energy integration range. 
 }
\end{figure}

\begin{figure}[t]
\includegraphics[scale=.4]{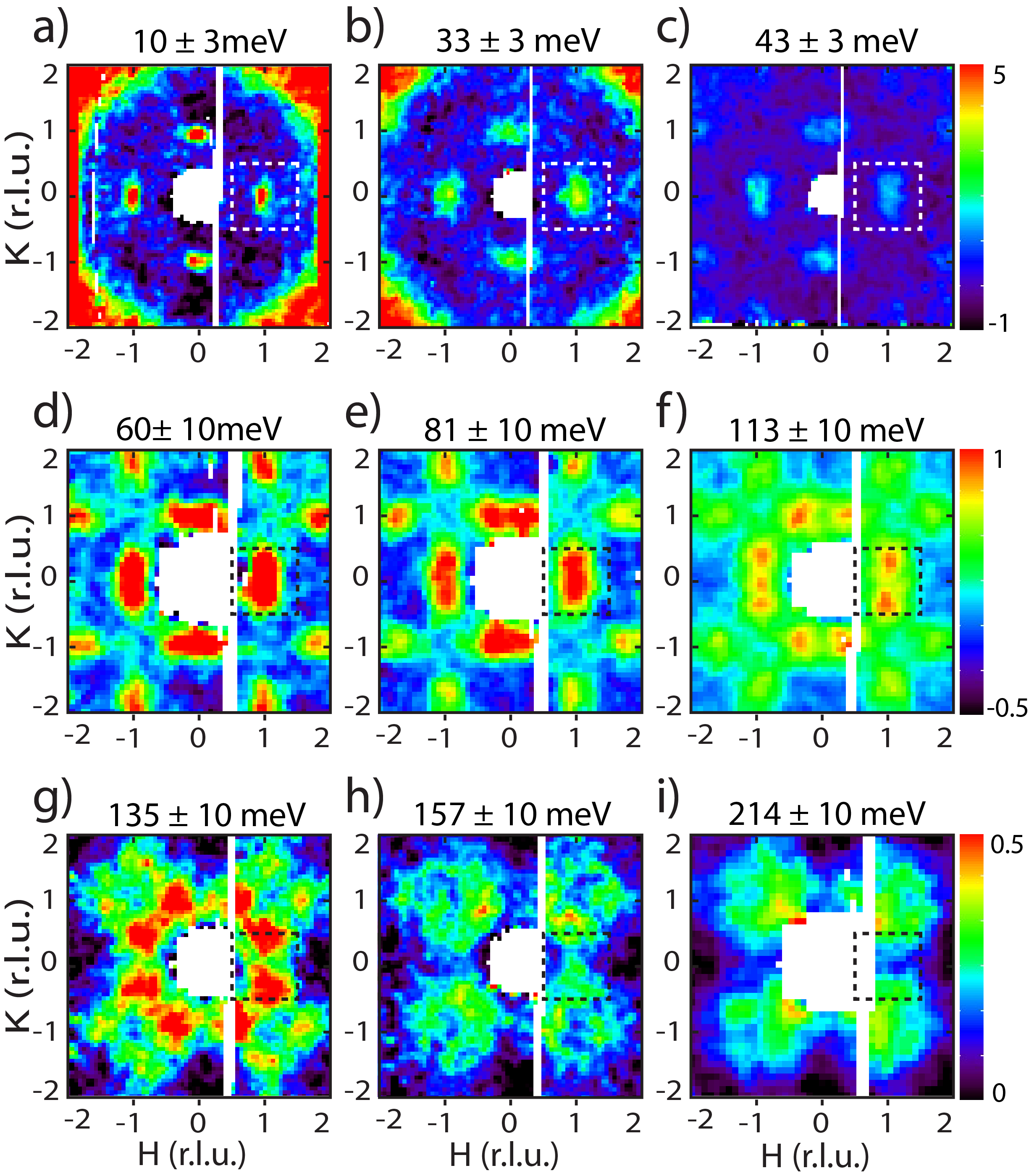}
\caption{
 Constant-energy slices through the magnetic excitations of BaFe$_{1.9}$Ni$_{0.1}$As$_{2}$ at different energies in several Brillouin zones.  The images were obtained after subtracting the background integrated from
$1.8<H<2.2$ and $-0.2<K<0.2$.  The color bars represent the vanadium normalized absolute spin excitation intensity in the units of
mbarn/sr/meV/f.u and the dashed boxes indicate AF zone boundaries 
for a single FeAs layer. Two dimensional images of spin excitations at
(a) $E=10\pm 3$, (b) $33\pm 3$, (c) $43\pm 3$, (d) $60\pm 10$, (e) $81\pm 10$, (f) $113\pm 10$,
(g) $135\pm 10$, (h) $157\pm 10$, and (i) $214\pm 10$ meV.
 }
\end{figure}

Figures 1c-e summarize our key findings for the electron-doped iron
arsenide superconductor BaFe$_{1.9}$Ni$_{0.1}$As$_{2}$ and the
comparison with the spin waves in BaFe$_2$As$_2$.  The data points in
Figs. 1c and 1d show the dispersion of spin excitations for optimal
doped BaFe$_{1.9}$Ni$_{0.1}$As$_{2}$ along $[1,K]$ and $[H,0]$, and 
the solid lines show the fit of BaFe$_2$As$_2$ spin waves 
to an effective Heisenberg $J_{1a}-J_{1b}-J_2$ model with $J_{1a}\neq J_{1b}$
\cite{harriger}.
Figure 1e shows the local dynamic susceptibility per formula unit
(f.u.), which contains two Fe(Ni) atoms, in absolute units, defined as
$\chi^{\prime\prime}(\omega)=\int{\chi^{\prime\prime}({\bf
    q},\omega)d{\bf q}}/\int{d{\bf q}}$ \cite{lester10},
where $\chi^{\prime\prime}({\bf q},\omega)=(1/3) tr( \chi_{\alpha \beta}^{\prime\prime}({\bf q},\omega))$,      
     at different
energies for BaFe$_2$As$_2$ and BaFe$_{1.9}$Ni$_{0.1}$As$_{2}$.  It is
clear that electron doping on BaFe$_2$As$_2$ only affects the
low-energy spin excitations by broadening the spin waves below
$80\,$meV, but has no impact on spin waves above 100 meV (see supplementary information).  The
quasiparticles that form within the spin-density-wave gap are
sensitive to the Fermi surface change upon doping BaFe$_2$As$_2$, and
hence the resulting low energy itinerant spin excitations
substantially change, while the higher energy spin excitations are
hardly affected.

\begin{figure}[t]
\includegraphics[scale=.37]{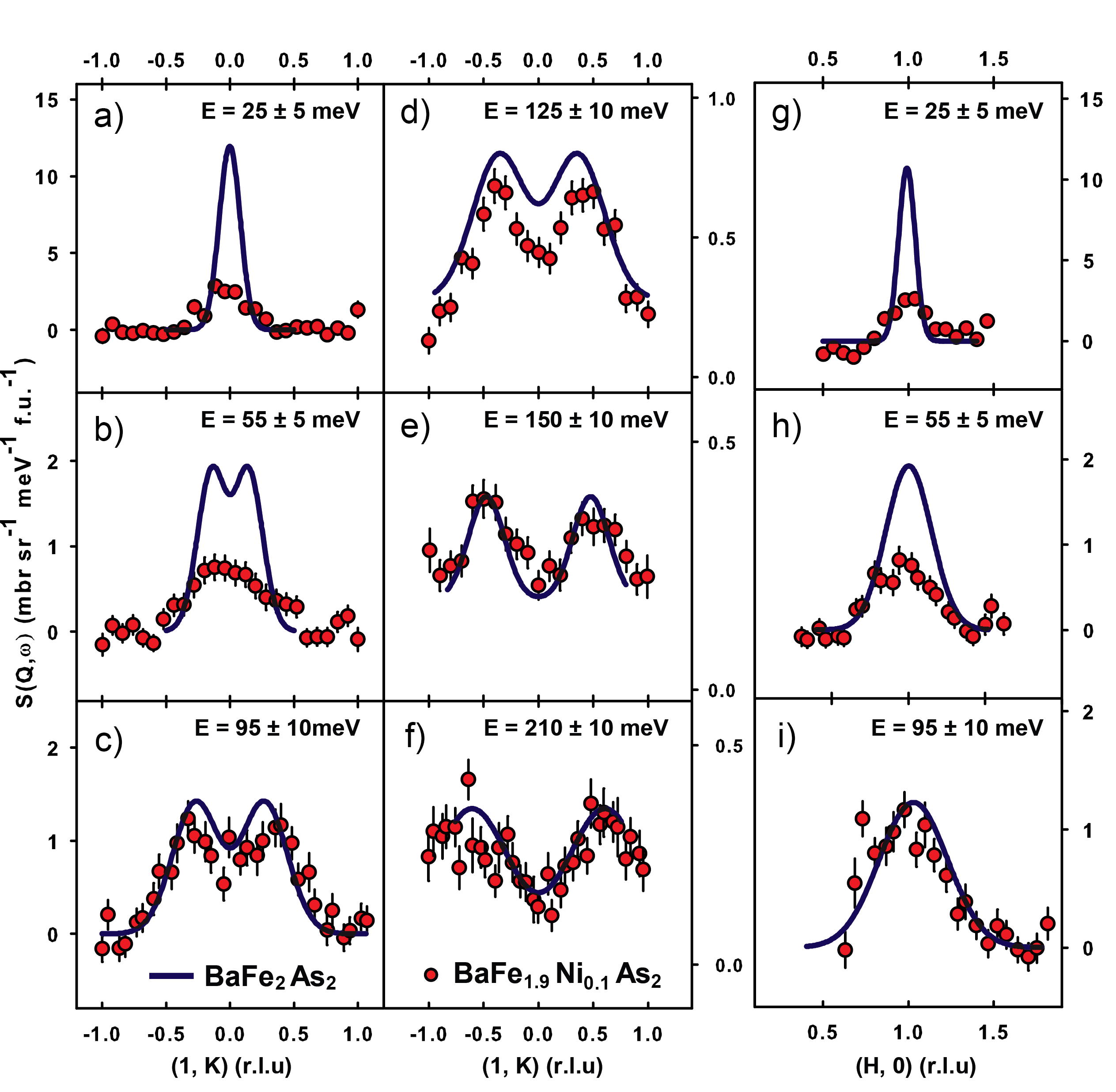}
\caption{
 Constant-energy cuts of the spin excitation dispersion as a function of
increasing energy along the $[1,K]$ and $[H,0]$ directions for BaFe$_{1.9}$Ni$_{0.1}$As$_{2}$.  The solid lines show
identical cuts for spin waves of BaFe$_2$As$_2$ in absolute units.
(a) Constant-energy cut along the $[1,K]$ direction at $E=25\pm 5$,
(b) $55\pm 5$, (c) $95\pm 10$, (d) $125\pm 10$, (e) $150\pm 10$, and (f) $210\pm 10$ meV.
(g) Constant-energy cut along the $[H,0]$ direction at $E=25\pm 5$, (h) $55\pm 5$, and $95\pm 10$ meV.
The error bars indicate the statistical errors of one standard deviation.
 }
\end{figure}

\begin{figure}[t]
\includegraphics[scale=.28]{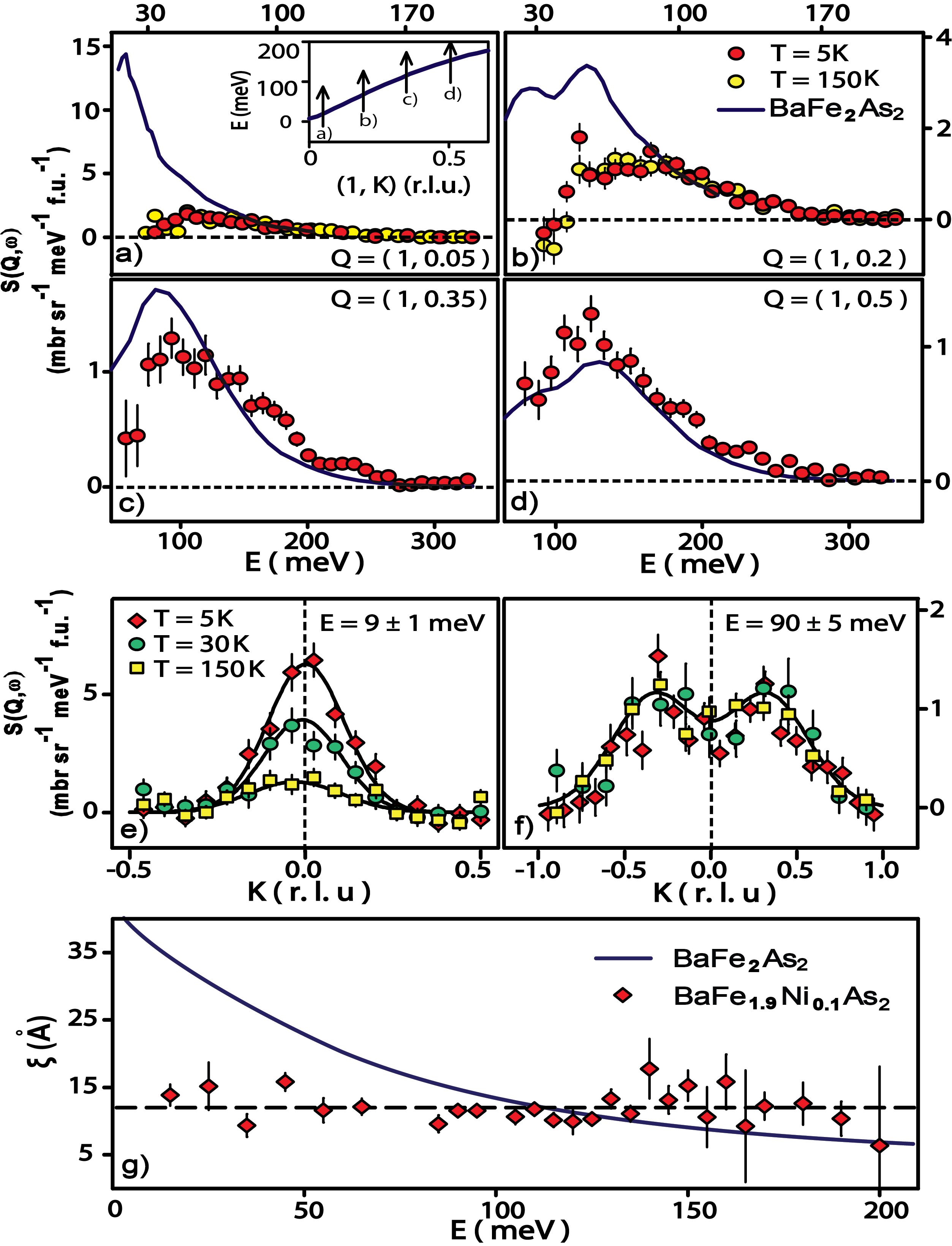}
\caption{
Constant-Energy/wave vector ($Q$) dependence of the spin excitations and dynamic spin-spin correlation lengths for
 BaFe$_{1.9}$Ni$_{0.1}$As$_{2}$ and BaFe$_2$As$_2$. (a)-(d) Constant-$Q$ cuts at $Q = (1, 0.05), (1, 0.2), (1, 0.35)$,
 and $(1, 0.5)$, respectively, at $T = 5$ (solid red circles) and 150 K (yellow filled circles) with background
 at $Q=(2,0)$ subtracted. The negative scattering in the data are due to over subtraction of the phonon background.
 The solid lines are identical cuts from spin waves
in BaFe$_2$As$_2$. For excitations below 100 meV, the intensity of the scattering of BaFe$_{1.9}$Ni$_{0.1}$As$_{2}$ is suppressed compared to
that of BaFe$_2$As$_2$.  For energies above 100 meV, the magnetic
scattering is virtually identical between the parent and superconductor.
(e) Constant-energy cuts at the neutron spin resonance energy of $E=9\pm 1$ meV \cite{chi} below
and above $T_c$.  The solid lines are Gaussian fits on
linear backgrounds. (f) Temperature dependence of spin excitations at $E=90\pm 5$ meV.
(g) Energy dependence of the dynamic spin-spin correlation lengths ($\xi$) at 5 K obtained by Fourier transform of constant-energy cuts
similar to those in Fig. 3a-f and Fig. 4e,f.
For all excitation energies probed ($10\leq E \leq 200$ meV), the dynamic spin-spin correlation lengths
are independent of energy. The solid line shows energy dependence of $\xi$ for BaFe$_2$As$_2$. The error bars indicate the statistical errors of one standard deviation.
 }
\end{figure}

To substantiate the key conclusions from the data and calculations presented in Figure 1, we show in Figure 2 the
two-dimensional constant-energy ($E$) images of
spin excitations of BaFe$_{1.9}$Ni$_{0.1}$As$_{2}$ in the
$(H, K)$ scattering plane for several Brillouin zones at 5 K.
In the undoped phase, spin waves in BaFe$_2$As$_2$ exhibits an anisotropy spin gap of $\Delta=9.8$ meV \cite{matan}.
On doping, the anisotropy spin gap disappears and spin excitations form transversely elongated ellipses that decrease in intensity with increasing energy \cite{lester10,hfli}.
For energy transfers of $E = 10\pm 3$ (Fig. 2a), $33\pm3$  (Fig. 2b), $43\pm 3$ (Fig. 2c),
$60\pm 10$  (Fig. 2d), and $81\pm 10$ meV (Fig. 2e), spin excitations are peaked at the AF wave vector $Q=(1,0)$
in the center of the Brillouin zone shown as dashed square boxes.
As the energy increases to $E = 113\pm 10$
(Fig. 2f) and $135\pm 10$ meV (Fig. 2g), spin excitations start to  split along the K-direction and form a ring
around the $\Gamma$ point.  Finally, spin excitations near the zone boundary at $E=157\pm 10$ and $214\pm 10$ meV
form four blobs centered at $Q=(1,1)$.

In order to determine the dispersion of spin excitations for BaFe$_{1.9}$Ni$_{0.1}$As$_{2}$,
we cut through the two-dimensional images similar to Fig. 2 along the $[1,K]$ and $[H,0]$ directions.
Figures 3a-3f show constant-energy cuts along the $[1,K]$ direction for $E=25\pm 5$, $55\pm 5$, $95\pm 10$,
$125\pm 10$, $150\pm 10$, and $210\pm 10$ meV.
The scattering becomes dispersive for spin excitation energies above 95 meV.  Figures 3g-3i show similar constant-energy
cuts along the $[H,0]$ direction. The solid lines in the Figure show identical spin wave cuts for BaFe$_2$As$_2$ \cite{harriger}.
Since both measurements were taken in absolute units, we can compare the impact of electron-doping on the spin waves in BaFe$_2$As$_2$.
At $E=25\pm 5$ meV, spin excitations in superconducting BaFe$_{1.9}$Ni$_{0.1}$As$_{2}$ are considerably broader in
momentum space and weaker in intensity than spin waves (Figs. 3a and 3g).  Upon increasing the
excitation energy to $55\pm 5$ meV, the dispersive spin waves in BaFe$_2$As$_2$ become weaker and broader (Figs. 3b and 3h). For energies above 95 meV, spin excitations in BaFe$_{1.9}$Ni$_{0.1}$As$_{2}$ are almost indistinguishable from spin waves
in BaFe$_2$As$_2$ in both the linewidths and intensity (Figs. 3c-3f, and 3i).  Based on these constant-energy cuts, we show in
Figs. 1c and 1d the comparison of spin excitation dispersions of BaFe$_{1.9}$Ni$_{0.1}$As$_{2}$ (filled circles and horizontal bars) with those of spin waves in BaFe$_2$As$_2$ (solid lines).
Inspection of Figs. 1-3 reveals that
electron-doping to BaFe$_2$As$_2$ only broadens and suppresses low energy
spin excitations and has no influence for spin waves above $100$ meV (see supplementary information).

To reveal further the effect of electron doping on spin waves of BaFe$_2$As$_2$,
we show in Figures 4a-4d constant-$Q$ cuts
at different wave vectors along the $[1,K]$ direction
for spin excitations in BaFe$_{1.9}$Ni$_{0.1}$As$_{2}$.  Near the Brillouin zone center at
$Q = (1,0.05)$ and $(1,0.2)$, well-defined spin excitations are observed near $E = 40$, and 60 meV
as shown in Figs. 4a and 4b, respectively.  The intensity of the scattering from the spin excitations in BaFe$_{1.9}$Ni$_{0.1}$As$_{2}$ is,
however, much lower than that of BaFe$_2$As$_2$ shown as solid lines in the Figures.
On increasing the wave vectors to $Q=(1,0.35)$ and $(1,0.5)$, the magnetic scattering
peak near $E = 100$, and 120 meV, and are essentially indistinguishable from spin waves in
BaFe$_2$As$_2$ as shown in Figs. 4c and 4d.  Furthermore, spin excitations have
virtually no temperature dependence between 5 K and 150 K (Fig. 4b).

Finally, we show in Figs. 4e and 4f temperature dependence of spin
excitations at energies near the neutron spin resonance $E=9$ meV
\cite{chi,mywang10} and at $E=90\pm 5$ meV, respectively. While the intensity
of the resonance at $E=9$ meV increases dramatically below $T_c$,
consistent with earlier work \cite{chi,mywang10}, spin excitations at $90\pm
5$ meV are identical on cooling from 150 K to 5 K.  We note that
high-energy spin waves in BaFe$_2$As$_2$ are also weakly temperature
dependent \cite{harriger}.  Figure 4g shows the energy dependence of the dynamic
spin-spin correlation lengths, which are about $\xi\approx 14$ \AA\
and excitation energy independent.  For comparison, the dynamic
spin-spin correlation length (the solid line in Fig. 4g) in
BaFe$_2$As$_2$ decreases with increasing energy and becomes similar to
that of BaFe$_{1.9}$Ni$_{0.1}$As$_{2}$ for excitation energies above
100 meV.

To check if spin excitations in the AF BaFe$_2$As$_2$ and
superconducting BaFe$_{1.9}$Ni$_{0.1}$As$_{2}$ can be understood in an
itinerant picture, we calculate the local susceptibility
$\chi^{\prime\prime}(\omega)$ using the random phase approximation
(RPA) based on realistic Fermi surfaces and band structures
\cite{park}.  Within RPA, the polarization bubble $\chi^0$ is computed
from the density function theory (DFT) Kohn-Sham Green's functions
while the irreducible vertex $\Gamma^{irr}$ is approximated by the
screened Coulomb parameters $\widetilde{U}$ and $\widetilde{J}$. Using
$\widetilde{U}=1.3\,$eV and $\widetilde{J}=0.4\,$eV and performing
calculations above $T_N$ \cite{park}, we find that the RPA estimated
$\chi^{\prime\prime}(\omega)$ for BaFe$_2$As$_2$ and
BaFe$_{1.9}$Ni$_{0.1}$As$_{2}$ (solid and dashed lines in Fig. 1b)
increases approximately linearly with energy and has absolute values
about a factor of three smaller than the observation (Fig. 1e).
Although the RPA calculation depends on Coulomb parameters used, we
note that the 5-orbital Hubbard model calculation using
$\widetilde{U}=0.8\,$eV and $\widetilde{J}=0.2\,$eV produces
essentially similar local magnetic spectra
\cite{graser}. Therefore, a pure RPA type itinerant model
underestimates the absolute spectral weight of the magnetic
excitations in iron pnictides.

The solid blue and red lines in Fig. 1b show the calculated local
susceptibility using a combined DFT and DMFT in the paramagnetic state.
Within DFT+DMFT, $\chi^{\prime\prime}({\bf q},\omega)$ is computed by
the Bethe-Salpeter equation using the polarization function $\chi^0$
and the two-particle local irreducible vertex function $\Gamma^{irr}$
\cite{park}.  $\chi^0$ is computed from the interacting one-particle
Green's function determined by the charge self-consistent full
potential DFT+DMFT method and $\Gamma^{irr}$ is extracted
from the two particle vertex function of the auxiliary impurity
problem. The latter is defined by the DMFT procedure using projection
of all electronic states to the $d$ character within the iron
muffin-tin sphere.
By comparing DFT+DMFT and RPA calculations in Fig.~1b
with data in Fig.~1e, we see that the former is much closer to the
observation.
Note that the calculation is done in the paramagnetic state, hence the
low energy modifications of the spectra due to the long range AF order
is not captured in this calculation.
RPA can describe only the itinerant part of the electron
spectra, while DFT+DMFT captures the essential aspects of both the
quasiparticles and the iron local moments formed by strong Hunds
coupling (see supplementary information for more detailed discussion). 
The improved agreement of DFT+DMFT thus suggest that both
the quasiparticles and the local moment aspects of the iron electrons
are needed to obtain the correct intensity and energy
distribution of neutron scattering spectra \cite{park}.

One way to quantitatively compare spin excitations in iron pnictides
with those in copper oxides is to estimate their total fluctuating
moments, defined as $\left\langle m^2\right\rangle=
(3\hbar/\pi)\int\chi^{\prime\prime}(\omega)d\omega/(1-\exp(-\hbar\omega/kT))$
\cite{lester10}.  Based on Fig. 1e, we find that $\left\langle
m^2\right\rangle=3.17\pm 0.16$ and $3.2\pm 0.16\ \mu_B^2$ per Fe(Ni) 
for BaFe$_2$As$_2$ and BaFe$_{1.9}$Ni$_{0.1}$As$_{2}$, respectively.
Using the formula for magnetic moment of a spin $\left\langle m^2\right\rangle= (g\mu_B)^2
S(S+1)$ (where $g=2$) \cite{lorenzana}, we find an effective iron spin $S$ of
about 1/2, similar to that of CaFe$_2$As$_2$ \cite{jzhao}. These
results also show that superconductivity in electron-doped system hardly
changes the total size of the fluctuating moment.  
In the fully localized (insulating) case,  the formal Fe$^{2+}$ oxidation state in BaFe$_2$As$_2$ would give a $3d^6$ electronic configuration. Hund's rules would yield $S=2$ and  $\left\langle
m^2\right\rangle=24\ \mu_B^2$ per Fe.  This is considerable more than the observed values suggesting that significant hybridization of Fe $3d$ with pnictide $p$ orbitals and among themselves, which leads to a metallic state where the Hund's coupling is less important than in the atomic limit \cite{Cvetkovic}.  For comparison, we
note that $\left\langle m^2\right\rangle > 1.9\ \mu_B^2$ per Cu for the AF insulating La$_2$CuO$_4$ measured over a similar energy range 
\cite{headings,hayden}.  From Fig. 1e, we see that the
large fluctuating moment $\left\langle m^2\right\rangle$ in iron
pnictides arises mostly from high-energy spin excitations that is
essentially temperature \cite{harriger} and electron-doping 
independent within the errors of our measurements (Fig. 1). 
Since there are currently no high-energy spin excitation data in absolute units for 
optimally hole-doped Ba$_{0.67}$K$_{0.33}$Fe$_2$As$_2$ \cite{clzhang}, it is unclear how hole-doping BaFe$_2$As$_2$ modifies the spin-wave spectra.

The DFT+DMFT calculation suggests that both the band structure and the local 
moment aspects (e.g. Hunds coupling) of the iron electrons are needed to obtain a good description of the magnetic response in
BaFe$_2$As$_2$ and BaFe$_{1.9}$Ni$_{0.1}$As$_{2}$. 
The weak electron-doping dependence of the fluctuating moment is consistent with the Hund's metal picture, where
electron filling associated with the Fe $3d^6$ electrons by Ni-doping is not expected to drastically affect the local moments. 
What is surprising is that the similarities in the local susceptibilities of the iron pnictides studied here and the parents of the cuprate superconductors.  The large fluctuating moment, arising from Hund's rule
coupling, and concentrated at higher energy in iron pnictides, nevertheless gives an
imprint on the massive and anisotropic low-energy
quasiparticles~\cite{Yin}, which form Cooper pairs at low energy.
This physics is different from the physics of doped charge transfer
insulators appropriate for copper oxides~\cite{palee}, hence the
electron correlations in iron pnictides and copper oxides have
different microscopic origins, although they are important for
understanding the magnetism and superconductivity for both materials.


\begin{flushleft}
{\bf Acknowledgements} 
We thank T. A. Maier, J. P. Hu, and Tao Xiang for helpful discussions.
The work at UTK is supported by the U.S. NSF-DMR-1063866 and 
NSF-OISE-0968226. Work at IOP is supported 
by the MOST of China 973 programs (2012CB821400, 2011CBA00110)
and NSFC-11004233.  
The work at Rutgers is supported by DOE BES DE-FG02-99ER45761 (GK),
ACS Petroleoum Research Fund 48802 and Alfred P. Sloan foundation (KH).

\end{flushleft}

\begin{flushleft}
{\bf Author contributions} 
P.D. and M.S.L. planned neutron scattering experiments.  M.S.L., L.W.H., H.Q.L., R.A.E., T.G., and P.D. carried out neutron scattering experiments.  Data analysis was done by M.S.L. with help from L.W.H., R.A.E., T.G., and S.M.H.  The samples were grown by H.Q.L. and co-aligned by M.S.L. and M.W.  The DFT and DMFT calculations were done by H.P., K.H., and G.K.  The paper was written by P.D., K.H., and G.K. with input from S.M.H. and M.S.L.  All coauthors provided comments on the paper.
\end{flushleft}

\begin{flushleft}
{\bf Additional information} 
The authors declare no competing financial interests. Supplementary information
accompanies this paper on www.nature.com/naturephysics. Reprints and permissions
information is available online at http://www.nature.com/reprints. Correspondence and
requests for materials should be addressed to P.D.
\end{flushleft}

\end{document}


\title{Nature of magnetic excitations in superconducting BaFe$_{1.9}$Ni$_{0.1}$As$_{2}$}
\author{Mengshu Liu}
\author{Leland W. Harriger}
\author{Huiqian Luo}
\author{Meng Wang}
\author{R. A. Ewings}
\author{T. Guidi}
\author{Hyowon Park}
\author{Kristjan Haule}
\author{Gabriel Kotliar}
\author{S. M. Hayden}
\author{Pengcheng Dai}

\maketitle

Our BaFe$_{2-x}$Ni$_{x}$As$_{2}$ electron-doped samples were grown using self flux method as described before \cite{yanchao}.  
Since the electronic and superconducting properties of Co and Ni-doping of BaFe$_2$As$_2$ near optimal 
superconducting transition temperatures are almost identical \cite{budko}, we chose to study spin excitations in optimally doped 
BaFe$_{1.9}$Ni$_{0.1}$As$_{2}$ with $T_c=20$ K.  To further compare spin excitations in BaFe$_2$As$_2$ and BaFe$_{1.9}$Ni$_{0.1}$As$_{2}$,
we show in Figure 1 spin excitations of these two materials at different energies.  Consistent with data shown in Figs. 2-4 of the main text, we see that the effect of electron-doping is to mostly modify spin excitations below 80 meV.
 
\begin{figure}
\includegraphics[scale=0.6]{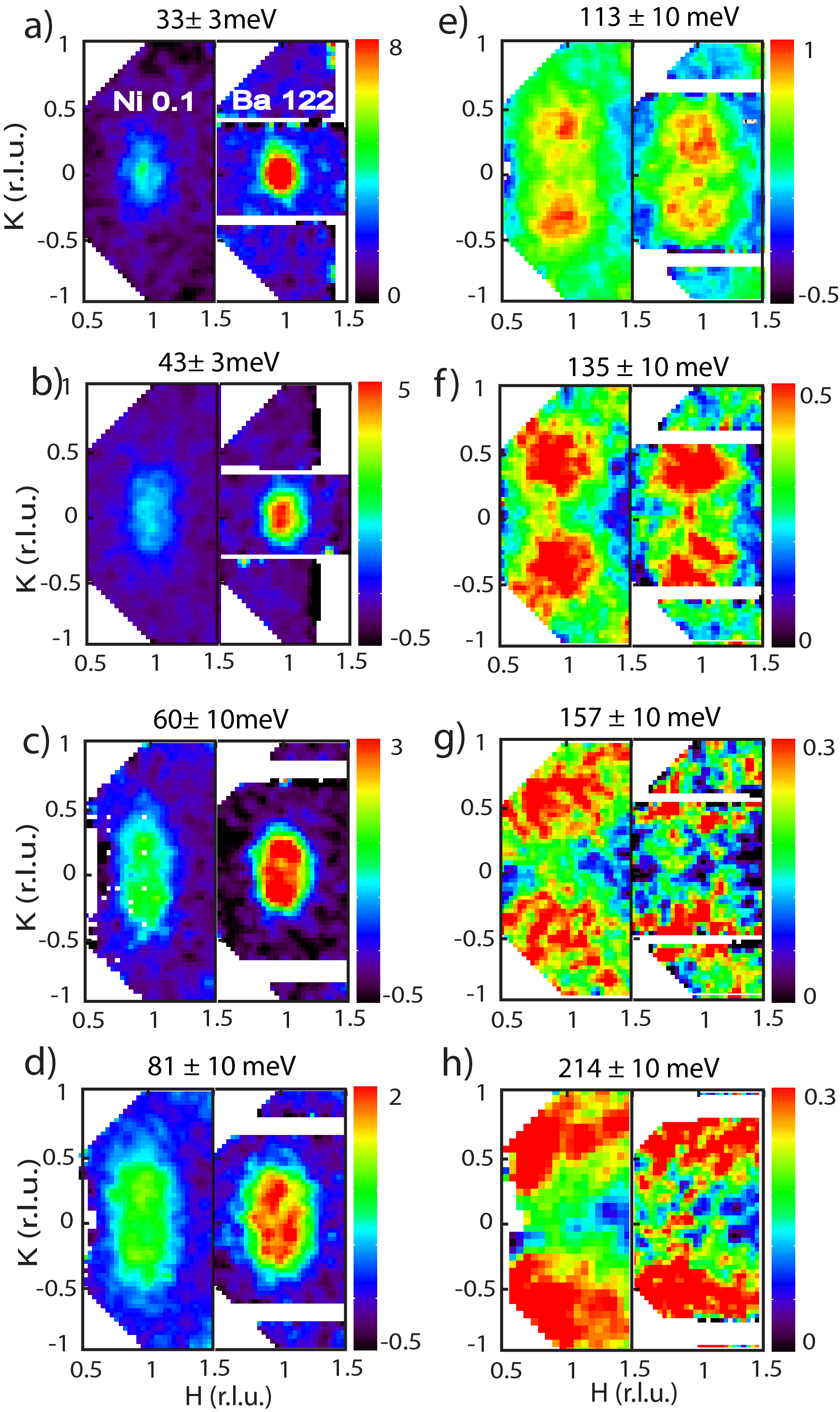}
\caption{ Constant-energy images of the spin excitations as a function of
increasing energy for BaFe$_{1.9}$Ni$_{0.1}$As$_{2}$ and
BaFe$_2$As$_2$ in units of mbarns/sr/meV/f.u.  
(a) $E=33\pm 3$, (b) $43\pm 3$, (c) $60\pm 10$, 
(d) $81\pm 10$, (e) $113\pm 10$, (f) $135\pm 10$, (g) $157\pm 10$, and (h) $214\pm 10$ meV.
}
\end{figure}

Our theoretical method for computing the magnetic excitation spectrum
employs abinitio full potential DFT+DMFT method, as implemented in
Ref. \cite{Haule}, which is based on the commercial DFT code of
Wien2k \cite{Wien2K}. The DMFT method requires solution of the
generalized quantum impurity problem, which is here solved by the
numerically exact continuous-time quantum Monte Carlo method
(CTQMC)~\cite{Haule:07,Werner:06}. The Coulomb interaction matrix for
electrons on iron atom was determined by the self-consistent GW method
in Ref.~\cite{Kutepov:10}, giving $U=5\,$eV and $J=0.8\,$eV for
the local basis functions within the all electron approach employed in
our DFT+DMFT method.

\begin{figure}
\includegraphics[scale=1]{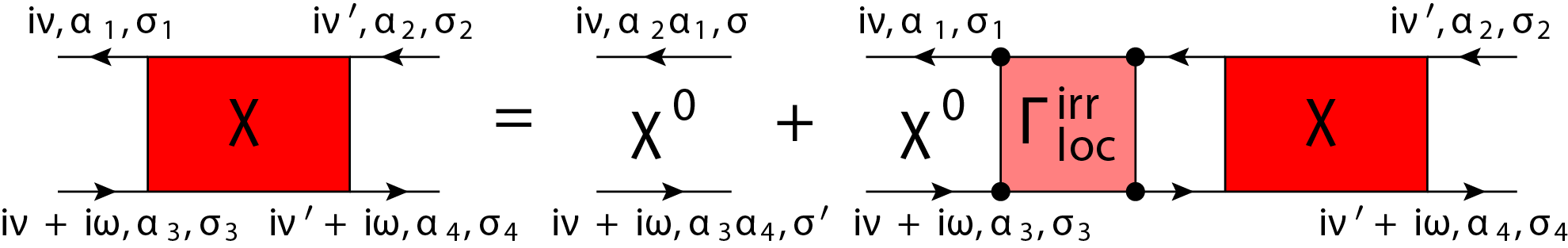}
\caption{ The Feynman diagrams for the Bethe-Salpeter equation.
  It relates the two-particle Green's function ($\chi$) with the
  polarization ($\chi^{0}$) and the local irreducible vertex function
  ($\Gamma_{loc}^{irr}$).   The non-local two-particle Green's
  function is obtained by replacing
  the local propagator by the non-local propagator.
  \label{fig:BSE}}
\end{figure}

The dynamical magnetic susceptibility $\chi(\textbf{q},\omega)$ is
computed from the \textit{ab initio} perspective by extracting the
two-particle vertex functions of DFT+DMFT solution
$\Gamma_{loc}^{irr}$. The polarization bubble $\chi^{0}$ is computed
from the fully interacting one particle Greens function.  The full
susceptibility is computed from $\chi^{0}$ and the two-particle
irreducible vertex function $\Gamma_{loc}^{irr}$, which is assumed to
be local in the same basis in which the DMFT self-energy is local,
implemented here by the projector to the muffin-tin
sphere~\cite{Haule}.  In order to extract $\Gamma_{loc}^{irr}$, we
employ the Bethe-Salpeter equation (see Fig.~\ref{fig:BSE}) which
relates the local two-particle Green's function ($\chi_{loc}$),
sampled by CTQMC, with both the local polarization function
($\chi_{loc}^{0}$) and $\Gamma_{loc}^{irr}$:
\begin{equation}
\Gamma_{loc{\alpha_{1}\sigma_{1},\alpha_{2}\sigma_{2}\atop
    \alpha_{3}\sigma_{3},\alpha_{4}\sigma_{4}}}^{irr}(i\nu,i\nu^{\prime})_{i\omega}=\frac{1}{T}[(\chi_{loc}^{0})_{i\omega}^{-1}-\chi_{loc}^{-1}].
\end{equation}
$\Gamma_{loc}^{irr}$ depends on three Matsubara frequencies ($i\nu$,
$i\nu^{\prime}$; $i\omega$), and both the spin ($\sigma_{1-4}$) and the orbital
($\alpha_{1-4}$) indices, which run over $3d$ states on the iron
atom. $T$ is the temperature.

Once the irreducible vertex $\Gamma_{loc}^{irr}$ is obtained, the
momentum dependent two-particle Green's function is constructed again
using the Bethe-Salpeter equation (Fig.~\ref{fig:BSE}) by replacing the
local polarization function $\chi_{loc}^{0}$ by the non-local one
$\chi_{\textbf{q},i\omega}^{0}$:
\begin{equation}
\chi_{{\alpha_{1}\sigma_{1},\alpha_{2}\sigma_{2}\atop
    \alpha_{3}\sigma_{3},\alpha_{4}\sigma_{4}}}(i\nu,i\nu^{\prime})_{\textbf{q},i\omega}=[(\chi^{0})_{\textbf{q},i\omega}^{-1}-T\cdot\Gamma_{loc}^{irr}]^{-1}.
\label{eq:BSE_imag}
\end{equation}
Finally, the dynamic
magnetic susceptibility $\chi(\textbf{q},i\omega)$ is obtained by
closing the two particle green's function with the magnetic moment
$\mu=\mu_B(\textbf{L}+2\textbf{S})$ vertex, and 
summing
over all internal degrees of freedom, i.e.,
orbitals ($\alpha_{1-4}$), spins ($\sigma_{1-4}$)
and frequencies ($i\nu$,$i\nu^{\prime}$),
on the four external legs
\begin{equation}
\chi(\textbf{q},i\omega)=T\sum_{i\nu,i\nu^{\prime}}\sum_{{\alpha_{1}\alpha_{2}\atop\alpha_{3}\alpha_{4}}}\sum_{{\sigma_{1}\sigma_{2}\atop
    \sigma_{3}\sigma_{4}}}\mu_{{\alpha_{1}\sigma_{1}\atop\alpha_{3}\sigma_{3}}}^{z}
\mu_{{\alpha_{2}\sigma_{2}\atop\alpha_{4}\sigma_{4}}}^{z} \;\chi_{{\alpha_{1}\sigma_{1},\alpha_{2}\sigma_{2}\atop
    \alpha_{3}\sigma_{3},\alpha_{4}\sigma_{4}}}(i\nu,i\nu^{\prime})_{\textbf{q},i\omega}
\label{eq:chi}
\end{equation}

We note that the same abinitio methodology which is here used to
compute the magnetic excitation spectra, was previously shown to
describe the photoemission, the optical spectra and the magnetic moments
of this material~\cite{Yin:11} in excellent agreement with experiment.